# Variationally consistent dynamics of nonlocal gradient elastic beams

Francesco P. Pinnola,

S. Ali Faghidian,

Raffaele Barretta,

Francesco Marotti de Sciarra

# Variationally consistent dynamics of nonlocal gradient elastic beams

F.P. Pinnola, S. Ali Faghidian, R. Barretta*, F. Marotti de Sciarra

*Department of Structures for Engineering and Architecture, University of Naples Federico II,*
*via Claudio 21, 80125 Naples, Italy*
*E-mails: francescopaolo.pinnola@unina.it - faghidian@gmail.com - rabarret@unina.it - marotti@unina.it*

## Abstract

The variational static formulation contributed in [International Journal of Engineering Science 143, 73-91 (2019)] is generalized in the present paper to model axial and flexural dynamic behaviors of elastic nano-beams by nonlocal strain and stress gradient approaches. Appropriate forms of non-standard boundary conditions are detected and properly prescribed. Equivalence between differential laws and integral convolutions is elucidated and invoked to analytically evaluate size-dependent axial and flexural fundamental frequencies of cantilever and fully-clamped beams which significantly characterize new-generation nano-actuators. The proposed methodology and ensuing results are tested by pertinent outcomes in literature. Advantageously, in comparison with available nonlocal gradient models, the developed formulation of elasticity leads to well-posed dynamic structural problems of nano-mechanics. Outcomes obtained by the well-established strain-driven and stress-driven nonlocal and gradient theories of Engineering Science are recovered under special ad hoc assumptions. The present study offers a simple and effective strategy to predict peculiar stiffening and softening dynamic responses of nano-scaled components of advanced technological devices, such as Nano-Electro-Mechanical-Systems (NEMS), and modern composite nano-structures.

*Corresponding author
E-mail: rabarret@unina.it

# 1. Introduction

Investigation of the mechanical response of nano-scaled structures and composites has gained significant attention due to their vast application in a broad spectrum of advanced nanotechnology including Nano-Electro-Mechanical-Systems (NEMS) (Marotti de Sciarra & Russo, 2019). Nano-composites are well-recognized to be superior compared to conventional materials (Almagableh et al., 2019). The outstanding properties of nano-composites, associated with privileged electrical (Wentzel & Sevostianov, 2018; Govorov et al., 2018) and thermal characteristics (Łydżba et al., 2019; Mazloum et al., 2019), are widely exploited for applications in modern NEMS. Beam-like structures are basic elements of NEMS devices, such as energy harvesters (Basutkar, 2019; Tran et al. 2018), piezoelectric actuators (Ouakad & Sedighi, 2019; Shirbani et al., 2018), actuated micro-plates (Medina et al., 2018; Farokhi & Ghayesh, 2018a), fluid conveying nano-tubes (Dehrouyeh-Semnani et al., 2019; Ghayesh et al., 2019a, 2019b) and nano-switches (SoltanRezaee & Afrashi, 2016). Local continuum mechanics is well-recognized to cease to hold in admissible description of mechanical behavior of nano-media. Thus, it is unavoidable to resort to generalized elasticity to significantly capture size-dependent structural responses. The thematic concerning with nanoscopic studies of field quantities is of major interest in literature of Engineering Science. Recent contributions are discussed as follows. Eringen's nonlocal elasticity has been employed to study size-dependent responses of nano-rods (Barretta et al., 2020a; Jalaei & Civalek, 2019; Numanoğlu & Civalek, 2019; Numanoğlu et al., 2018), nano-beams (She et al., 2017; Fernández-Sáez & Zaera, 2017; Zhu et al., 2017; Demir & Civalek, 2017; Fernández-Sáez et al., 2016) and nano-plates (Srividhya et al., 2018). Axial behavior of nano-rods (Barretta et al., 2019a), flexure of nano-beams (Kim & Zeidi, 2018; Khakalo et al., 2018; Ghayesh, 2018, 2019; Lurie & Solyaev, 2018; Qi et al., 2018; Jiao & Alavi, 2018; Attia & Abdel Rahman, 2018) and nano-plates (Farokhi & Ghayesh, 2018b; Farokhi et al., 2017) have been also investigated via different forms of strain gradient theory. In nonlocal strain gradient elasticity, axial responses of nano-rods (Zhu & Li, 2017a, 2017b), torsion of FG nano-beams (Barretta et al., 2020b), linear and nonlinear flexure of nano-beams (Farajpour et al., 2019; Ghayesh & Farajpour, 2018; Li et al., 2018; Faghidian, 2018a; She et al., 2018a, 2018b; Lu et al., 2017), size-dependent mechanics of nano-plates (Lu et al., 2018) and elastodynamic responses of nano-shells (Faleh et al., 2018) have been furthermore examined. Extra results can be found in (Ghayesh & Farajpour, 2019; Farajpour et al., 2018).

In the framework of nonlocal elasticity, exploited by Eringen (1983) for unbounded domains, the integral convolution between elastic strain and a nonlocal kernel is adopted for simulation of nonlocal long-range interactions. However, Eringen's strain-driven nonlocal integral model clashes against serious obstructions as applied to structures of technical interest, since the stress field associated with the nonlocal constitutive law does not meet the requirements dictated by equilibrium (Romano et al., 2018). A well-posed theory of nonlocal elasticity was introduced by Romano & Barretta (2017) via the stress-driven formulation, where input and output fields of the integral convolution of Eringen's nonlocal theory are commuted. Extension of the stress-driven nonlocal elasticity to two phase models (Barretta et al., 2018a, 2019b), curved nano-beams (Barretta et al., 2019c), functionally graded materials (Barretta et al., 2018b), thermo-elasticity (Barretta et al., 2018c) and two-dimensional problems (Barretta et al., 2019d) has been also addressed and discussed. Nonlocal and gradient continuum formulations of elasticity theory can be merged exploiting different thermodynamic frameworks (Lim et al., 2015; Faghidian, 2018b, 2018c). The resulted nonlocal strain gradient constitutive model is of higher-order than the one of the classical local problem, and therefore, requires suitable non-standard boundary conditions.

A preliminary answer to this crucial issue in Engineering Science was provided by Barretta & Marotti de Sciarra (2018) and exploited in (Apuzzo et al., 2018, 2019), revealing the need of prescribing constitutive boundary conditions to properly close the relevant nonlocal problem. The appropriate choice of non-standard boundary conditions, affecting particularly the mechanical behaviour of nano-structures has been in dispute in literature (Zaera et al., 2019). A decisive response to this issue was given in (Barretta & Marotti de Sciarra, 2019) by developing the proper variational formulation of the elasto-static problem of nonlocal strain and stress gradient of inflected beams that, unlike alleged claims in literature, is well-posed. Advantageously, both softening and stiffening size-dependent structural responses have been shown to be predictable with the new nonlocal methodology which provides also most of elasticity theories exploited by the scientific community to capture scale phenomena, such as modified nonlocal strain gradient model, strain- and stress-driven local/nonlocal elasticity.

## 2. Motivation and outline

Elastodynamic analysis of beam-type structural components of NEMS is a topic of major interest in Engineering Science. Available outcomes on the matter are obtained in literature without taking the proper non-standard constitutive boundary conditions into due account. The motivation of the present paper is to extend and apply the variationally consistent nonlocal gradient continuum mechanics, conceived by Barretta & Marotti de Sciarra (2019) for static problems, to free vibrations of elastic nano-beams. The plan is as follows. Preliminary assumptions and equations governing local beams are recalled in Sect. 3. Sect. 4 is devoted to the development of nonlocal strain and stress gradient formulations of axial and flexural free vibrations of slender beams. Fundamental frequencies are analytically detected for cantilever and fully-clamped beams in Sect. 5 where numerical results are tabulated, illustrated, compared and commented upon. Closing remarks are outlined in Sect. 6.

## 3. Local elastic beams

A homogeneous straight beam of length $L$, with uniform cross-section $\Xi$ and material density $\rho$ is considered. The abscissa $x$ is taken along the beam axis and the pair of cross-sectional Cartesian coordinates is denoted by $y$ and $z$, with $z$ flexural axis. To formulate the local beam model, the elasticity solution of Saint-Venant's flexure problem (Romano et al., 2012; Faghidian, 2016) can be exploited. The beam kinematics is described in terms of $x$- and $z$-coordinates. The displacement filed $\mathbf{u}$, based on Bernoulli-Euler theory, along the coordinate directions is expressed by

$$u_1 = u(x,t) - z\,\partial_x w(x,t), \quad u_2 = 0, \quad u_3 = w(x,t) \tag{1}$$

where $u, w : [0, L] \mapsto \Re$ represent axial and transverse displacements at time $t$. The shear deformation is disregarded in the formulation of Bernoulli-Euler model (Faghidian, 2017), and accordingly, the only non-vanishing strain along the beam axis is the longitudinal component that writes as

$$\varepsilon_{xx} = \varepsilon(x,t) - z\,\kappa(x,t) = \partial_x u(x,t) - z\,\partial_{xx} w(x,t) \tag{2}$$

with $\varepsilon : [0, L] \to \Re$ and $\kappa : [0, L] \to \Re$ extension and curvature of the beam centroidal axis.

Cross-sectional mass $A_\rho$ and rotatory inertia $I_\rho$, to be utilized in the differential conditions of dynamic equilibrium, are recalled as

$$A_\rho = \iint_\Xi \rho(z)\,dA, \qquad I_\rho = \iint_\Xi \rho(z) z^2 dA \tag{3}$$

The flexure of the beam is attributed to an equilibrated force system $\mathbf{F}$ composed of distributed axial and transverse loadings $p : [0, L] \to \Re$ and $q : [0, L] \to \Re$.

The equilibrated stresses in a Bernoulli–Euler beam are then an axial force field $N:[0,L]\to\Re$ and a bending moment field $M:[0,L]\to\Re$. Equilibrium is imposed by the virtual work condition

$$\langle \mathbf{F},\delta\mathbf{u}\rangle = \int_0^L N\,\delta\varepsilon\,dx + \int_0^L M\,\delta\kappa\,dx \qquad (4)$$

with

$$\langle \mathbf{F},\delta\mathbf{u}\rangle = \int_0^L \left(p - A_\rho\partial_{tt}u\right)\delta u\,dx + \int_0^L \left(q - A_\rho\partial_{tt}w + I_\rho\partial_{xxtt}w\right)\delta w\,dx \qquad (5)$$

being the virtual work of the force system $\mathbf{F}$, for any axial and transversal virtual displacement fields $\delta u,\delta w:[0,L]\to\Re$ fulfilling homogeneous kinematic boundary conditions. The scalar functions $\delta\varepsilon := \partial_x\delta u:[0,L]\to\Re$ and $\delta\kappa := \partial_{xx}\delta w:[0,L]\to\Re$ are the kinematically compatible extension and flexural curvature associated with axial and transversal virtual displacements $\delta u,\delta w:[0,L]\to\Re$. In consequence of applying a standard localization procedure, while performing integration by parts, the differential and standard boundary conditions associated with the aforementioned variational scheme are provided by

$$
\begin{aligned}
&\partial_x N + p = A_\rho\partial_{tt}u\\
&\partial_{xx}M = q - A_\rho\partial_{tt}w + I_\rho\partial_{xxtt}w\\
&N\,\delta u\big|_{x=0} = N\,\delta u\big|_{x=L} = 0\\
&M\left(\partial_x\delta w\right)\big|_{x=0} = M\left(\partial_x\delta w\right)\big|_{x=L} = 0\\
&\left(\partial_x M\right)\delta w\big|_{x=0} = \left(\partial_x M\right)\delta w\big|_{x=L} = 0
\end{aligned}
\qquad (6)
$$

Note that, kinematics and statics of a Bernoulli–Euler beam have been briefly recalled above without making references to any constitutive aspects which will be introduced hereafter.

## 4. Nonlocal gradient elastic nano-beams

The nonlocal gradient theory of elasticity has been recently formulated in variational terms by Barretta & Marotti de Sciarra (2019) for nano-beams subjected to static flexure. The consistent variational formulation leads to well-posed structural problems of nano-mechanics. The nonlocal gradient elasticity is extended in the present paper to investigate both axial and flexural elastodynamic responses of nano-beams. Both Nonlocal strain-driven Gradient (NstrainG) and Nonlocal stress-driven Gradient (NstressG) theories of elasticity are developed by making recourse to a variationally consistent constitutive formulation.

Let us preliminarily recall the definition of integral convolution between a nonlocal kernel $\psi_c : \Re \to \,]0,\infty[$ and a scalar field $f$

$$\left(\psi_c * f\right)(x) := \int_0^L \psi_c\left(x - \zeta\right) f\left(\zeta\right) d\zeta \tag{7}$$

with $x, \zeta$ points of the structural domain $[0, L]$ and $c \in \,]0,\infty[$ nonlocal length-scale parameter. The averaging kernel $\psi_c$ is selected in such a way to fulfill positivity and symmetry, normalization and impulsivity properties (Romano et al., 2018).

In elastodynamic analysis of nano-beams, it is convenient to introduce axial and flexural local elastic stiffnesses $A_E, I_E$ as

$$A_E = \iint_{\Xi} E\left(z\right) dA, \qquad I_E = \iint_{\Xi} E\left(z\right) z^2 dA \tag{8}$$

where $E$ is Euler-Young elastic modulus.

## 4.1. *Nonlocal strain-driven gradient elasticity*

Motivated by the seminal approach of Barretta & Marotti de Sciarra (2019), the nonlocal strain-driven gradient (NstrainG) model for axial and flexural deformation of elastic nano-beams is governed by the potential $\Pi_{\text{NstrainG}}$

$$\begin{aligned}
\Pi_{\text{NstrainG}}\left(\varepsilon,\kappa\right) &:= \frac{1}{2}A_E\int_0^L\left[\alpha\varepsilon^2+\left(1-\alpha\right)\left(\psi_c*\varepsilon\right)\varepsilon+\ell^2\left(\psi_c*\partial_x\varepsilon\right)\partial_x\varepsilon\right]dx \\
&\quad +\frac{1}{2}I_E\int_0^L\left[\alpha\kappa^2+\left(1-\alpha\right)\left(\psi_c*\kappa\right)\kappa+\ell^2\left(\psi_c*\partial_x\kappa\right)\partial_x\kappa\right]dx
\end{aligned} \tag{9}$$

with $\alpha\in\left[0,1\right]$ being a mixture parameter along with $\ell\in\left[0,\infty\right[$ representing a gradient length-scale parameter. The nonlocal gradient axial field $N$ and bending moment field $M$ associated with NstrainG is provided by the variational condition

$$\begin{aligned}
\left\langle N,\delta\varepsilon\right\rangle &:= \int_0^L N\left(x,t\right)\delta\varepsilon\left(x,t\right)dx = \left\langle d\,\Pi_{\text{NstrainG}}\left(\varepsilon,\kappa\right),\delta\varepsilon\right\rangle \\
\left\langle M,\delta\kappa\right\rangle &:= \int_0^L M\left(x,t\right)\delta\kappa\left(x,t\right)dx = \left\langle d\,\Pi_{\text{NstrainG}}\left(\varepsilon,\kappa\right),\delta\kappa\right\rangle
\end{aligned} \tag{10}$$

for any virtual extension field $\delta\varepsilon\in C_0^1\left(\left[0,L\right];\Re\right)$ and flexural curvature field $\delta\kappa\in C_0^1\left(\left[0,L\right];\Re\right)$ having compact supports in the beam domain. The NstrainG constitutive law can be determined via evaluation of the directional derivatives of the elastic strain energy $\Pi_{\text{NstrainG}}$ along the virtual extension and flexural curvature fields while integrating by parts

$$\left\langle d\,\Pi_{\text{NstrainG}}\left(\varepsilon,\kappa\right),\delta\varepsilon\right\rangle = A_E \int_0^L \left[\alpha\varepsilon\delta\varepsilon + \left(1-\alpha\right)\left(\psi_c * \varepsilon\right)\delta\varepsilon + \ell^2\left(\psi_c * \partial_x\varepsilon\right)\left(\partial_x\delta\varepsilon\right)\right]dx$$

$$= A_E \int_0^L \left[\alpha\varepsilon + \left(1-\alpha\right)\left(\psi_c * \varepsilon\right) - \ell^2\partial_x\left(\psi_c * \partial_x\varepsilon\right)\right]\delta\varepsilon\,dx$$

$$+ A_E \ell^2\left[\left(\psi_c * \partial_x\varepsilon\right)\delta\varepsilon\big|_{x=L} - \left(\psi_c * \partial_x\varepsilon\right)\delta\varepsilon\big|_{x=0}\right]$$

$$\left\langle d\,\Pi_{\text{NstrainG}}\left(\varepsilon,\kappa\right),\delta\kappa\right\rangle = I_E \int_0^L \left[\alpha\kappa\delta\kappa + \left(1-\alpha\right)\left(\psi_c * \kappa\right)\delta\kappa + \ell^2\left(\psi_c * \partial_x\kappa\right)\left(\partial_x\delta\kappa\right)\right]dx \qquad (11)$$

$$= I_E \int_0^L \left[\alpha\kappa + \left(1-\alpha\right)\left(\psi_c * \kappa\right) - \ell^2\partial_x\left(\psi_c * \partial_x\kappa\right)\right]\delta\kappa\,dx$$

$$+ I_E \ell^2\left[\left(\psi_c * \partial_x\kappa\right)\delta\kappa\big|_{x=L} - \left(\psi_c * \partial_x\kappa\right)\delta\kappa\big|_{x=0}\right]$$

Both virtual kinematic fields $\delta\varepsilon$ and $\delta\kappa$ have compact supports, i.e. $\delta\varepsilon\big|_{x=L} = \delta\varepsilon\big|_{x=0} = 0$ and $\delta\kappa\big|_{x=L} = \delta\kappa\big|_{x=0} = 0$, and the boundary terms in Eq. (11) are therefore vanished. The explicit expression of the sought nonlocal gradient constitutive relation is determined by prescribing the variational condition Eq. (10) upon localization

$$N\left(x,t\right) = A_E \left[\alpha\varepsilon\left(x,t\right) + \left(1-\alpha\right)\left(\psi_c * \varepsilon\right)\left(x,t\right) - \ell^2\partial_x\left(\psi_c * \partial_x\varepsilon\right)\left(x,t\right)\right]$$
$$M\left(x,t\right) = I_E \left[\alpha\kappa\left(x,t\right) + \left(1-\alpha\right)\left(\psi_c * \kappa\right)\left(x,t\right) - \ell^2\partial_x\left(\psi_c * \partial_x\kappa\right)\left(x,t\right)\right] \qquad (12)$$

A suitable choice for the nonlocal kernel $\psi_c$ is the bi-exponential averaging function, generally adapted in nonlocal elasticity, and well-established to meet the required positivity, symmetry, normalization and impulsivity requirements (Romano et al., 2018)

$$\psi_c\left(x\right) := \frac{1}{2c}\exp\left(-\frac{|x|}{c}\right) \qquad (13)$$

Following the strategy exploited in the Prop. 3.1 of the contribution by Barretta & Marotti de Sciarra (2019), the equivalent differential constitutive law of NstrainG integro-differential law equipped with suitable constitutive boundary conditions is established.

**Proposition 1. NstrainG constitutive equivalency**

The NstrainG constitutive laws for elastic nano-beams Eq. (12), equipped with the bi-exponential kernel Eq. (13) defined on a bounded interval $[0, L]$ are equivalent to the differential relations

$$\frac{1}{c^2} N(x,t) - \partial_{xx} N(x,t) = \frac{1}{c^2} A_E \varepsilon(x,t) - A_E \left( \alpha + \frac{\ell^2}{c^2} \right) \partial_{xx} \varepsilon(x,t)$$

$$\frac{1}{c^2} M(x,t) - \partial_{xx} M(x,t) = \frac{1}{c^2} I_E \kappa(x,t) - I_E \left( \alpha + \frac{\ell^2}{c^2} \right) \partial_{xx} \kappa(x,t)$$

(14)

subjected to the following four constitutive boundary conditions (CBCs) at beam ends

$$\partial_x N(0,t) - \frac{1}{c} N(0,t) = -A_E \frac{\alpha}{c} \varepsilon(0,t) + A_E \left( \alpha + \frac{\ell^2}{c^2} \right) \partial_x \varepsilon(0,t)$$

$$\partial_x N(L,t) + \frac{1}{c} N(L,t) = A_E \frac{\alpha}{c} \varepsilon(L,t) + A_E \left( \alpha + \frac{\ell^2}{c^2} \right) \partial_x \varepsilon(L,t)$$

$$\partial_x M(0,t) - \frac{1}{c} M(0,t) = -I_E \frac{\alpha}{c} \kappa(0,t) + I_E \left( \alpha + \frac{\ell^2}{c^2} \right) \partial_x \kappa(0,t)$$

$$\partial_x M(L,t) + \frac{1}{c} M(L,t) = I_E \frac{\alpha}{c} \kappa(L,t) + I_E \left( \alpha + \frac{\ell^2}{c^2} \right) \partial_x \kappa(L,t)$$

(15)

The following well-established special cases adapted in mechanics of nano-structures, being of strain-driven kind, can be recovered by NstrainG.

The modified nonlocal strain gradient constitutive law equipped with suitable CBCs can be recovered as the mixture parameter vanishes $\alpha \to 0$ (Apuzzo et al., 2018)

$$\frac{1}{c^2}N(x,t) - \partial_{xx}N(x,t) = \frac{1}{c^2}A_E\,\varepsilon(x,t) - A_E\frac{\ell^2}{c^2}\partial_{xx}\varepsilon(x,t)$$

$$\frac{1}{c^2}M(x,t) - \partial_{xx}M(x,t) = \frac{1}{c^2}I_E\,\kappa(x,t) - I_E\frac{\ell^2}{c^2}\partial_{xx}\kappa(x,t)$$

$$\partial_x N(0,t) - \frac{1}{c}N(0,t) = A_E\frac{\ell^2}{c^2}\partial_x\varepsilon(0,t)$$

$$\partial_x N(L,t) + \frac{1}{c}N(L,t) = A_E\frac{\ell^2}{c^2}\partial_x\varepsilon(L,t) \qquad (16)$$

$$\partial_x M(0,t) - \frac{1}{c}M(0,t) = I_E\frac{\ell^2}{c^2}\partial_x\kappa(0,t)$$

$$\partial_x M(L,t) + \frac{1}{c}M(L,t) = I_E\frac{\ell^2}{c^2}\partial_x\kappa(L,t)$$

Similarly, the two-phase local/nonlocal strain-driven model and associated CBCs can be detected as the gradient characteristic length approaches zero $\ell \to 0$ (Fernández-Sáez & Zaera, 2017)

$$\frac{1}{c^2}N(x,t) - \partial_{xx}N(x,t) = \frac{1}{c^2}A_E\,\varepsilon(x,t) - \alpha A_E\partial_{xx}\varepsilon(x,t)$$

$$\frac{1}{c^2}M(x,t) - \partial_{xx}M(x,t) = \frac{1}{c^2}I_E\,\kappa(x,t) - \alpha I_E\partial_{xx}\kappa(x,t)$$

$$\partial_x N(0,t) - \frac{1}{c}N(0,t) = -A_E\frac{\alpha}{c}\varepsilon(0,t) + \alpha A_E\partial_x\varepsilon(0,t)$$

$$\partial_x N(L,t) + \frac{1}{c}N(L,t) = A_E\frac{\alpha}{c}\varepsilon(L,t) + \alpha A_E\partial_x\varepsilon(L,t) \qquad (17)$$

$$\partial_x M(0,t) - \frac{1}{c}M(0,t) = -I_E\frac{\alpha}{c}\kappa(0,t) + \alpha I_E\partial_x\kappa(0,t)$$

$$\partial_x M(L,t) + \frac{1}{c}M(L,t) = I_E\frac{\alpha}{c}\kappa(L,t) + \alpha I_E\partial_x\kappa(L,t)$$

### 4.2. *Nonlocal stress-driven gradient elasticity*

The original idea of the stress-driven model of nonlocal elasticity (Romano & Barretta, 2017) consists in exchanging the roles of the nonlocal and source fields of the nonlocal integral convolution. However, the strain- and stress-driven elasticity theories cannot be inferred to be the inverse of one another (Romano et al., 2018). Indeed, strain- and stress-driven constitutive laws represent two distinct models based on the physical interpretation of nonlocal and source fields. The elastic potential of nano-beam $\Pi_{\text{NstressG}}$ consistent with the NstressG is accordingly introduced as

$$
\begin{aligned}
\Pi_{\text{NstressG}}\left(N,M\right) &:= \frac{1}{2}\frac{1}{A_E}\int_0^L\left[\alpha N^2+(1-\alpha)\left(\psi_c*N\right)N+\ell^2\left(\psi_c*\partial_x N\right)\partial_x N\right]dx \\
&+\frac{1}{2}\frac{1}{I_E}\int_0^L\left[\alpha M^2+(1-\alpha)\left(\psi_c*M\right)M+\ell^2\left(\psi_c*\partial_x M\right)\partial_x M\right]dx
\end{aligned}
\tag{18}
$$

In the framework of NstressG, the extension $\varepsilon$ and curvature $\kappa$ of the elastic nano-beam is provided by the following variational conditions

$$
\begin{aligned}
\left\langle\varepsilon,\delta N\right\rangle &:= \int_0^L\varepsilon\left(x,t\right)\delta N\left(x,t\right)dx=\left\langle d\Pi_{\text{NstressG}}\left(N,M\right),\delta N\right\rangle \\
\left\langle\kappa,\delta M\right\rangle &:= \int_0^L\kappa\left(x,t\right)\delta M\left(x,t\right)dx=\left\langle d\Pi_{\text{NstressG}}\left(N,M\right),\delta M\right\rangle
\end{aligned}
\tag{19}
$$

for any virtual axial force field $\delta N\in C_0^1\left(\left[0,L\right];\Re\right)$ and a bending moment field $\delta M\in C_0^1\left(\left[0,L\right];\Re\right)$ having compact supports in the structural domain. Evaluation of the directional derivative of the elastic potential $\Pi_{\text{NstressG}}$ while performing integration by parts results in

$$
\begin{aligned}
\left\langle d\Pi_{\text{NstressG}}\left(N,M\right),\delta N\right\rangle &= \frac{1}{A_E}\int_0^L\left[\alpha N\,\delta N+(1-\alpha)\left(\psi_c*N\right)\delta N+\ell^2\left(\psi_c*\partial_x N\right)\left(\partial_x\delta N\right)\right]dx \\
&= \frac{1}{A_E}\int_0^L\left[\alpha N+(1-\alpha)\left(\psi_c*N\right)-\ell^2\partial_x\left(\psi_c*\partial_x N\right)\right]\delta N\,dx \\
&+\frac{1}{A_E}\ell^2\left[\left(\psi_c*\partial_x N\right)\delta N\big|_{x=L}-\left(\psi_c*\partial_x N\right)\delta N\big|_{x=0}\right] \\
\left\langle d\Pi_{\text{NstressG}}\left(N,M\right),\delta M\right\rangle &= \frac{1}{I_E}\int_0^L\left[\alpha M\,\delta M+(1-\alpha)\left(\psi_c*M\right)\delta M+\ell^2\left(\psi_c*\partial_x M\right)\left(\partial_x\delta M\right)\right]dx \\
&= \frac{1}{I_E}\int_0^L\left[\alpha M+(1-\alpha)\left(\psi_c*M\right)-\ell^2\partial_x\left(\psi_c*\partial_x M\right)\right]\delta M\,dx \\
&+\frac{1}{I_E}\ell^2\left[\left(\psi_c*\partial_x M\right)\delta M\big|_{x=L}-\left(\psi_c*\partial_x M\right)\delta M\big|_{x=0}\right]
\end{aligned}
\tag{20}
$$

Since the virtual kinetic fields $\delta N$ and $\delta M$ are supposed to have compact supports in the structural domain, the boundary terms in Eq. (20) are disappearing.

The nonlocal gradient constitutive law of the extension $\varepsilon$ and curvature $\kappa$ of the nano-beam can be detected via imposing the variational condition Eq. (19) and applying a standard localization procedure

$$\varepsilon(x,t) = \frac{1}{A_E}\Big[\alpha N(x,t) + (1-\alpha)(\psi_c * N)(x,t) - \ell^2 \partial_x (\psi_c * \partial_x N)(x,t)\Big]$$

$$\kappa(x,t) = \frac{1}{I_E}\Big[\alpha M(x,t) + (1-\alpha)(\psi_c * M)(x,t) - \ell^2 \partial_x (\psi_c * \partial_x M)(x,t)\Big]$$

(21)

Likewise, assuming the nonlocal kernel to be the bi-exponential function, the equivalent differential constitutive law and the consequent CBCs consistent with NstressG model can be established.

**Proposition 2. NstressG constitutive equivalency**

The nonlocal gradient constitutive relation Eq. (21), endowed with the bi-exponential kernel Eq. (13), is equivalent to the differential constitutive law

$$\frac{1}{c^2}\varepsilon(x,t) - \partial_{xx}\varepsilon(x,t) = \frac{1}{c^2}\frac{1}{A_E}N(x,t) - \frac{1}{A_E}\left(\alpha + \frac{\ell^2}{c^2}\right)\partial_{xx}N(x,t)$$

$$\frac{1}{c^2}\kappa(x,t) - \partial_{xx}\kappa(x,t) = \frac{1}{c^2}\frac{1}{I_E}M(x,t) - \frac{1}{I_E}\left(\alpha + \frac{\ell^2}{c^2}\right)\partial_{xx}M(x,t)$$

(22)

equipped with the constitutive boundary conditions (CBC)

$$\partial_x \varepsilon(0,t) - \frac{1}{c}\varepsilon(0,t) = -\frac{1}{A_E}\frac{\alpha}{c}N(0,t) + \frac{1}{A_E}\left(\alpha + \frac{\ell^2}{c^2}\right)\partial_x N(0,t)$$

$$\partial_x \varepsilon(L,t) + \frac{1}{c}\varepsilon(L,t) = \frac{1}{A_E}\frac{\alpha}{c}N(L,t) + \frac{1}{A_E}\left(\alpha + \frac{\ell^2}{c^2}\right)\partial_x N(L,t)$$

$$\partial_x \kappa(0,t) - \frac{1}{c}\kappa(0,t) = -\frac{1}{I_E}\frac{\alpha}{c}M(0,t) + \frac{1}{I_E}\left(\alpha + \frac{\ell^2}{c^2}\right)\partial_x M(0,t)$$

$$\partial_x \kappa(L,t) + \frac{1}{c}\kappa(L,t) = \frac{1}{I_E}\frac{\alpha}{c}M(L,t) + \frac{1}{I_E}\left(\alpha + \frac{\ell^2}{c^2}\right)\partial_x M(L,t)$$

(23)

The well-recognized particular size-dependent models of elastic nano-beams, being of stress-driven kind, are also included in the NstressG elasticity.

Setting the gradient characteristic length zero $\ell \to 0$, the two-phase local/nonlocal stress-driven model and associated CBCs are recovered (Barretta et al., 2018d)

$$\frac{1}{c^2}\varepsilon(x,t) - \partial_{xx}\varepsilon(x,t) = \frac{1}{c^2}\frac{1}{A_E}N(x,t) - \frac{\alpha}{A_E}\partial_{xx}N(x,t)$$

$$\frac{1}{c^2}\kappa(x,t) - \partial_{xx}\kappa(x,t) = \frac{1}{c^2}\frac{1}{I_E}M(x,t) - \frac{\alpha}{I_E}\partial_{xx}M(x,t)$$

$$\partial_x\varepsilon(0,t) - \frac{1}{c}\varepsilon(0,t) = -\frac{1}{A_E}\frac{\alpha}{c}N(0,t) + \frac{\alpha}{A_E}\partial_x N(0,t)$$

$$\partial_x\varepsilon(L,t) + \frac{1}{c}\varepsilon(L,t) = \frac{1}{A_E}\frac{\alpha}{c}N(L,t) + \frac{\alpha}{A_E}\partial_x N(L,t) \qquad (24)$$

$$\partial_x\kappa(0,t) - \frac{1}{c}\kappa(0,t) = -\frac{1}{I_E}\frac{\alpha}{c}M(0,t) + \frac{\alpha}{I_E}\partial_x M(0,t)$$

$$\partial_x\kappa(L,t) + \frac{1}{c}\kappa(L,t) = \frac{1}{I_E}\frac{\alpha}{c}M(L,t) + \frac{\alpha}{I_E}\partial_x M(L,t)$$

The well-posed stress-driven nonlocal model of inflected nano-beams conceived by Romano & Barretta (2017) can be also demonstrated to be a particular case of the NstressG via vanishing the mixture and gradient characteristic parameters $\alpha, \ell \to 0$

$$\frac{1}{c^2}\varepsilon(x,t) - \partial_{xx}\varepsilon(x,t) = \frac{1}{c^2}\frac{1}{A_E}N(x,t)$$

$$\frac{1}{c^2}\kappa(x,t) - \partial_{xx}\kappa(x,t) = \frac{1}{c^2}\frac{1}{I_E}M(x,t)$$

$$\partial_x\varepsilon(0,t) - \frac{1}{c}\varepsilon(0,t) = 0$$

$$\partial_x\varepsilon(L,t) + \frac{1}{c}\varepsilon(L,t) = 0 \qquad (25)$$

$$\partial_x\kappa(0,t) - \frac{1}{c}\kappa(0,t) = 0$$

$$\partial_x\kappa(L,t) + \frac{1}{c}\kappa(L,t) = 0$$

## 5. Free vibration analysis

Variationally consistent nonlocal strain- and stress-driven gradient models of elasticity are adapted to investigate free vibration response of structural schemes of nano-mechanics interest: cantilever and fully-clamped Bernoulli-Euler nano-beams. The non-dimensional terms: gyration radius $\bar{r}$, nonlocal and gradient characteristic parameters $\lambda_c$ and $\lambda_\ell$, axial and flexural fundamental frequencies $\bar{\Omega}$ and $\bar{\omega}$ are introduced as

$$\bar{r} = \frac{1}{L}\sqrt{\frac{I_\rho}{A_\rho}}, \qquad \lambda_c = \frac{c}{L}, \qquad \lambda_\ell = \frac{\ell}{L}, \qquad \bar{\Omega}^2 = \frac{L^2 A_\rho}{\pi^2 A_E}\Omega^2, \qquad \bar{\omega}^2 = \frac{L^4 A_\rho}{I_E}\omega^2 \qquad (26)$$

### 5.1. *Axial free vibrations*

To investigate the axial free vibrations of elastic nano-beams, the distributed axial loading $p$ is allowed to vanish. Prescribing the differential condition of equilibrium Eq. (6)$_1$ to the constitutive differential law of NstrainG Eq. (14)$_1$ or NstressG Eq. (22)$_1$ results in the expression of the axial force filed $N$ as

$$\frac{1}{c^2}N_{\text{NstrainG}}(x,t) = A_\rho \partial_{tt}\varepsilon(x,t) + \frac{1}{c^2}A_E\varepsilon(x,t) - A_E\left(\alpha + \frac{\ell^2}{c^2}\right)\partial_{xx}\varepsilon(x,t)$$

$$\frac{1}{c^2}N_{\text{NstressG}}(x,t) = \left(\alpha + \frac{\ell^2}{c^2}\right)A_\rho \partial_{tt}\varepsilon(x,t) + \frac{1}{c^2}A_E\varepsilon(x,t) - A_E\partial_{xx}\varepsilon(x,t) \qquad (27)$$

Applying the axial kinematic compatibility, the differential condition of dynamic equilibrium governing axial vibrations of nano-beams associated with the NstrainG writes thus as

$$\frac{1}{c^2}A_E\partial_{xx}u(x,t) - A_E\left(\alpha + \frac{\ell^2}{c^2}\right)\partial_{xxxx}u(x,t) = \frac{1}{c^2}A_\rho \partial_{tt}u(x,t) - A_\rho \partial_{ttxx}u(x,t) \qquad (28)$$

and consistent with NstressG is given by

$$\frac{1}{c^2}A_E\partial_{xx}u(x,t)-A_E\partial_{xxxx}u(x,t)=\frac{1}{c^2}A_\rho\partial_{tt}u(x,t)-\left(\alpha+\frac{\ell^2}{c^2}\right)A_\rho\partial_{ttxx}u(x,t)\tag{29}$$

subject to the standard boundary conditions Eq. (6)$_3$ and corresponding constitutive boundary conditions associated with NstrainG Eq. (15)$_{1,2}$ or NstressG Eq. (23)$_{1,2}$. A standard procedure of separating spatial and time variables is applied to analyze axial free vibrations

$$u(x,t)=U(x)\exp(i\Omega t)\tag{30}$$

with $i=\sqrt{-1}$, $U$ and $\Omega$ denoting the spatial mode shapes and natural frequency of axial vibrations. Imposing the separation of variables Eq. (30) on the differential conditions of dynamic equilibrium Eq. (28-29), the differential condition of axial spatial mode shapes for NstrainG

$$-A_E\left(\alpha+\frac{\ell^2}{c^2}\right)\frac{d^4U}{dx^4}(x)+\left(\frac{1}{c^2}A_E-A_\rho\Omega^2\right)\frac{d^2U}{dx^2}(x)+\frac{1}{c^2}A_\rho\Omega^2U(x)=0\tag{31}$$

and NstressG can be found as

$$-A_E\frac{d^4U}{dx^4}(x)+\left(\frac{1}{c^2}A_E-\left(\alpha+\frac{\ell^2}{c^2}\right)A_\rho\Omega^2\right)\frac{d^2U}{dx^2}(x)+\frac{1}{c^2}A_\rho\Omega^2U(x)=0\tag{32}$$

The axial spatial mode shapes can be analytically detected as

$$U(x)=\sum_{k=1}^{4}U_k\exp(\xi_k x)\tag{33}$$

where unknown integration constants $U_k\ (k=1..4)$ yet needed to be determined along with $\xi_k\ (k=1..4)$ being the roots of the characteristic equation associated with the differential equations of (31) or (32).

The solution procedure for determianion of the axial natural frequencies of elastic nano-beams is throughly discussed in (Barretta et al., 2019e). Accrodingly, prescription of the standard boundary conditions Eq. (6)₃ along with the constitutive boundary conditions consistent with NstrainG Eq. (15)₁,₂ or NstressG Eq. (23)₁,₂ to the axial spatial mode shapes Eq. (33) will result in a homogeneous fourth-order algebraic system in terms of the unknown constants $U_k$ $(k = 1..4)$. It is well-establsihed that the system of algebraic equations has to be singular in order to detect the non-trivial solution of axial free vibrations, and thus, the determinant of the coefficients of the homogeneous fourth-order system has to vanish. A highly nonlinear characteristic equation in terms of axial natural frequency is detected for NstrainG and NstressG elastic nano-beams with cantilever and fully-clamped boundary conditions that is numerically solved.

Figs. 1-4 demonstrate the fundamental axial frequencies of cantilever and fully-clamped nano-beams consistent with the NstrainG and NstressG elasticity models. The detected axial frequencies are furthermore normalized applying the corresponding local natural frequencies $\bar{\Omega}_{\text{LOC}}$. In Figs. 1-4, the nonlocal and gradient characteristic parameters $\lambda_c, \lambda_\ell$ are respectively ranging in the intervals $]0,1[$ and $]0.1,1[$, while two values of the mixture parameter as $\alpha = 0$ and $\alpha = 1/2$ are prescribed.

It is deduced from the illustrative results consistent with the NstrainG model that the nonlocal characteristic parameter $\lambda_c$ has the effect of decreasing the fundamental axial frequency, i.e. a larger $\lambda_c$ involves a smaller natural axial frequency. The natural axial frequencies consistent with the NstrainG model accordingly reveal a softening response in terms of the nonlocal characteristic parameter. The axial frequencies consistent with the NstrainG model also increase by increasing the gradient characteristic $\lambda_\ell$ or the mixture parameter $\alpha$, thus demonstrating a stiffening response in terms of the gradient and the mixture parameters. On the contrary, a softening behavior is detected for the axial frequencies associated with NstressG for increasing the gradient or mixture parameter and a stiffening response is confirmed for increasing nonlocal characteristic parameter. As the small-scale characteristic and mixture parameters approach zero, fundamental axial frequencies of the local elastic beam can be recovered. Numerical values of normalized fundamental axial frequencies of cantilever and fully-clamped nano-beams detected in accordance with NstrainG and NstressG models are collected in Tables 1 through 4, for different values of the characteristic and mixture parameters.

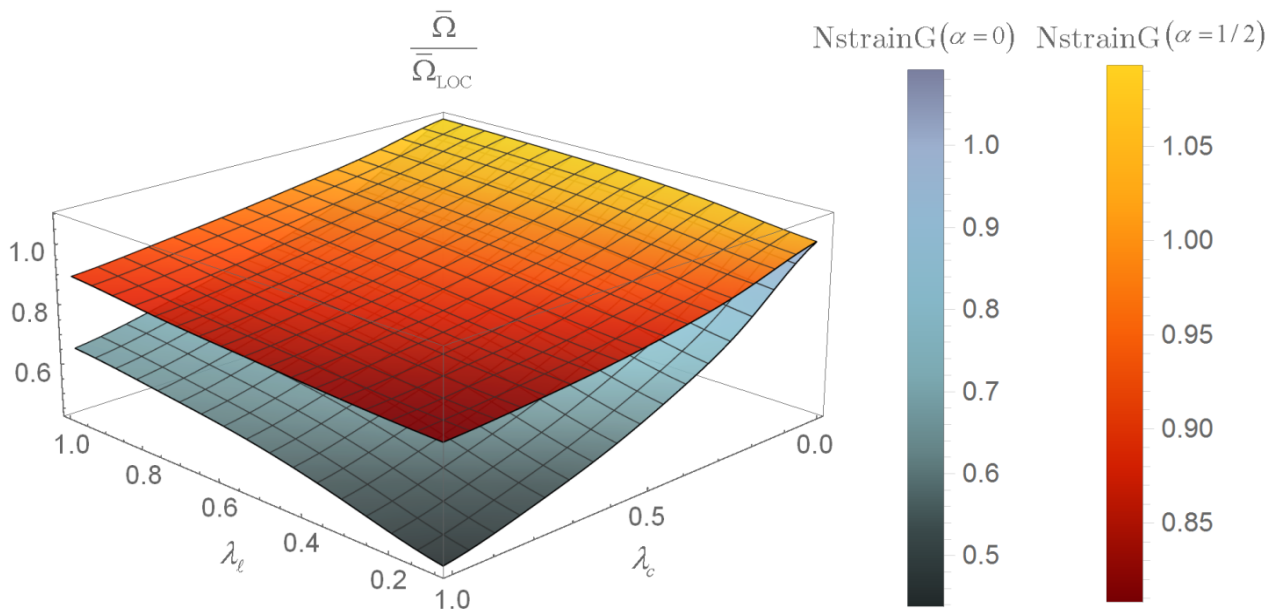

**Fig. 1.** Normalized axial fundamental frequency of NstrainG cantilever nano-beams

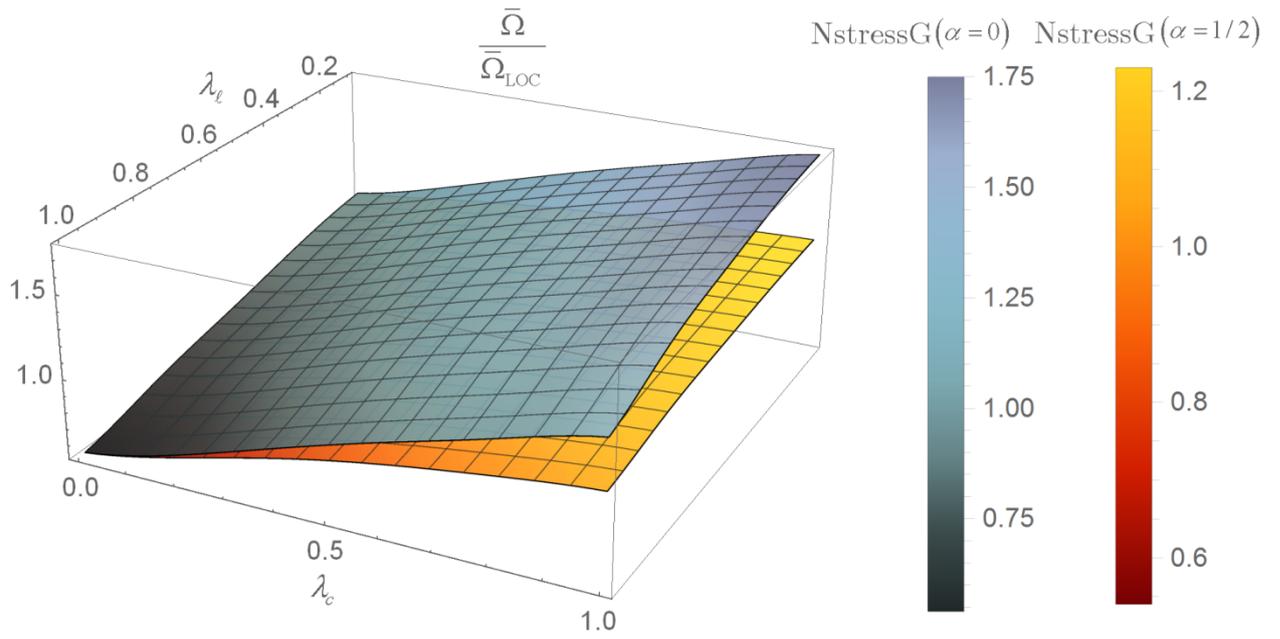

**Fig. 2.** Normalized axial fundamental frequency of NstressG cantilever nano-beams

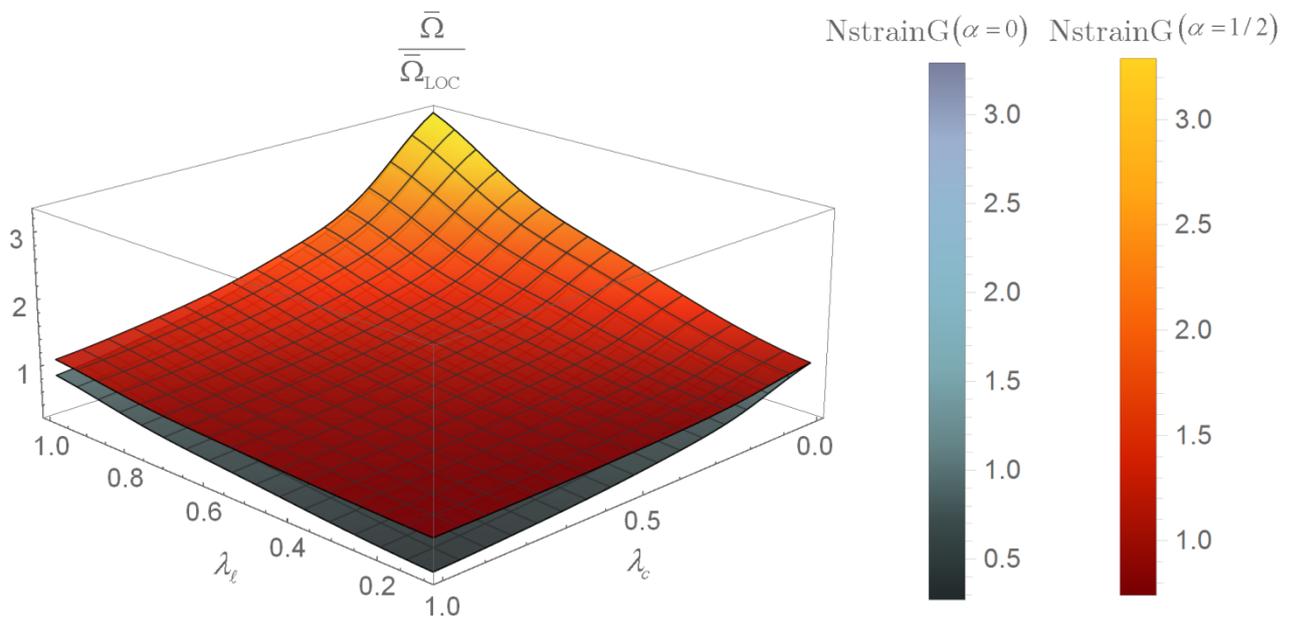

**Fig. 3.** Normalized axial fundamental frequency of NstrainG fully-clamped nano-beams

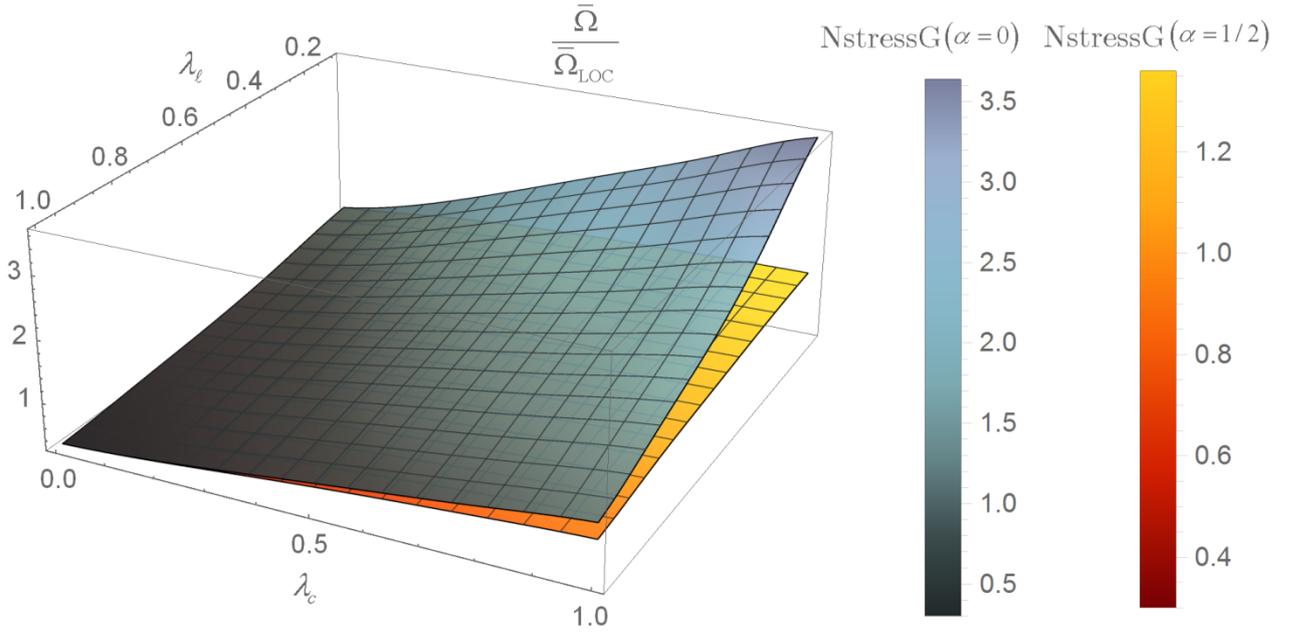

**Fig. 4.** Normalized axial fundamental frequency of NstressG fully-clamped nano-beams

## 5.2. *Flexural free vibrations*

To examine the flexural free vibrations of elastic nano-beams, distributed transverse loading $q$ is assumed to vanish in the differential condition of equilibrium. The bending moment field $M$ is detected by applying the differential condition of equilibrium Eq. (6)$_2$ to the constitutive differential law of NstrainG Eq. (14)$_2$ or NstressG Eq. (22)$_2$ as

$$\frac{1}{c^2}M_{\text{NstrainG}}(x,t) = -A_\rho \partial_{tt}w + I_\rho \partial_{xxtt}w + \frac{1}{c^2}I_E \kappa(x,t) - I_E\left(\alpha + \frac{\ell^2}{c^2}\right)\partial_{xx}\kappa(x,t)$$

$$\frac{1}{c^2}M_{\text{NstressG}}(x,t) = \frac{1}{c^2}I_E\kappa(x,t) - I_E\partial_{xx}\kappa(x,t) + \left(\alpha + \frac{\ell^2}{c^2}\right)\left(-A_\rho\partial_{tt}w + I_\rho\partial_{xxtt}w\right)$$

(34)

Employing the flexural kinematic compatibility, the differential condition of dynamic equilibrium governing the flexural vibrations of NstrainG nano-beams is given by

$$\partial_{xx}\left(-A_\rho\partial_{tt}w+I_\rho\partial_{xxtt}w\right)+\partial_{xx}\left(\frac{1}{c^2}I_E\partial_{xx}w\left(x,t\right)\right)-\partial_{xx}\left(I_E\left(\alpha+\frac{\ell^2}{c^2}\right)\partial_{xxxx}w\left(x,t\right)\right)$$
$$=\frac{1}{c^2}\left(-A_\rho\partial_{tt}w+I_\rho\partial_{xxtt}w\right)$$

(35)

and for NstressG nano-beams writes as

$$\partial_{xx}\left(\frac{1}{c^2}I_E\partial_{xx}w\left(x,t\right)\right)-\partial_{xx}\left(I_E\partial_{xxxx}w\left(x,t\right)\right)+\partial_{xx}\left(\left(\alpha+\frac{\ell^2}{c^2}\right)\left(-A_\rho\partial_{tt}w+I_\rho\partial_{xxtt}w\right)\right)$$
$$=\frac{1}{c^2}\left(-A_\rho\partial_{tt}w+I_\rho\partial_{xxtt}w\right)$$

(36)

equipped with the standard boundary conditions Eq. (6)$_{4,5}$ and corresponding constitutive boundary conditions consistent with NstrainG Eq. (15)$_{3,4}$ or NstressG Eq. (23)$_{3,4}$. In the same way, natural frequencies and mode shapes of flexural vibrations are evaluated by employing again the classical separation of spatial and time variables as

$$w\left(x,t\right)=W\left(x\right)\exp\left(i\,\omega t\right)$$

(37)

with $\omega$ designating the natural frequency of flexural vibrations. Enforcing the separation of variables Eq. (37) to the differential condition of dynamic equilibrium Eq. (35-36), the governing equation on the flexural spatial mode shape $W$ is obtained for NstrainG model as

$$-I_E\left(\alpha+\frac{\ell^2}{c^2}\right)\frac{d^6W}{dx^6}\left(x\right)+\left(\frac{1}{c^2}I_E-I_\rho\omega^2\right)\frac{d^4W}{dx^4}\left(x\right)+\left(A_\rho\omega^2+\frac{1}{c^2}I_\rho\omega^2\right)\frac{d^2W}{dx^2}\left(x\right)$$
$$-\frac{1}{c^2}A_\rho\omega^2W\left(x\right)=0$$

(38)

and detected for NstressG model as

$$-I_E\frac{d^6W}{dx^6}\left(x\right)+\left(\frac{1}{c^2}I_E-\left(\alpha+\frac{\ell^2}{c^2}\right)I_\rho\omega^2\right)\frac{d^4W}{dx^4}\left(x\right)+\left(\left(\alpha+\frac{\ell^2}{c^2}\right)\left(A_\rho\omega^2\right)+\frac{1}{c^2}I_\rho\omega^2\right)\frac{d^2W}{dx^2}\left(x\right)$$
$$-\frac{1}{c^2}A_\rho\omega^2W\left(x\right)=0$$

(39)

The analytical solution of the aforementioned governing equations (38-39) of the flexural spatial mode shape can be expressed by (Barretta et al., 2018b)

$$W\left(x\right)=\sum_{k=1}^{6}W_{k}\exp\left(\eta_{k}x\right) \tag{40}$$

where $W_{k}\left(k=1..6\right)$ are unknown integration constants desired to be determined by standard and constitutive boundary conditions, along with $\eta_{k}\left(k=1..6\right)$ being the roots of the characteristic equation associated with the differential equations of (38) or (39).

A homogeneous sixth-order algebraic system in terms of the unknown constants $W_{k}\left(k=1..6\right)$ is similarly formulated for the flexural vibrations, by adopting the solution form of the flexural spatial mode shape Eq. (40) and imposing the standard boundary conditions Eq. (6)$_{4,5}$ together with the constitutive boundary conditions associated with NstrainG Eq. (15)$_{3,4}$ or NstressG Eq. (23)$_{3,4}$. In order to obtain a non-trivial solution of flexural free vibrations, the determinant of the coefficients of the homogeneous sixth-order system should also vanish. The solution procedure results in a highly nonlinear characteristic equation for either of NstrainG or NstressG elasticity models, in terms of the fundamental flexural frequency, which is also numerically solved.

The effects of characteristic and mixture parameters on the fundamental flexural frequency associated with the NstrainG and NstressG elasticity models are exhibited in Figs. 5-8 for cantilever and fully-clamped nano-beams. While the characteristic and mixture parameters are assumed to have the same ranging set as the axial free vibrations, exhibited in Figs. 1 through 4, the non-dimensional gyration radius of the nano-beam is given by $\bar{r}=1/20$. Once more, the illustrated fundamental flexural frequencies are normalized utilizing the corresponding local counterparts of the Bernoulli-Euler beam model $\bar{\omega}_{\mathrm{LOC}}$.

As noticeably inferred from the numerical illustrations associated with the NstrainG model, while the flexural fundamental frequencies exhibit a stiffening behavior in terms of the gradient and the mixture parameters, a softening behavior in terms of the nonlocal characteristic parameter is detected for both cantilever and fully-clamped nano-beams. In contrast, a softening response is demonstrated in the framework of NstressG for increasing the gradient or mixture parameter and a stiffening behavior is detected for increasing nonlocal characteristic parameter. The fundamental flexural frequency of nano-beam in accordance with either of the NstrainG or NstressG models coincides with the fundamental flexural frequency of local beam theory for vanishing small-scale characteristic and mixture parameters. The normalized fundamental flexural frequencies of cantilever and fully-clamped nano-beams, associated with NstrainG and NstressG models, are numerically evaluated and collected in Tables 5 through 8 for several values of the characteristic and mixture parameters.

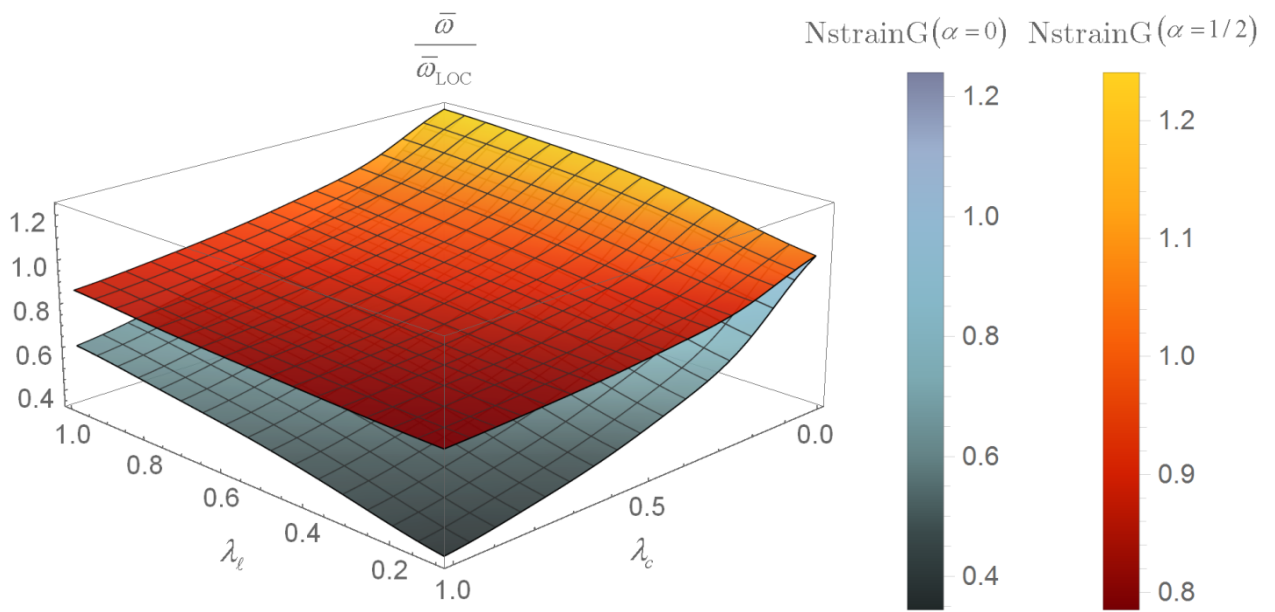

**Fig. 5.** Normalized flexural fundamental frequency of NstrainG cantilever nano-beams

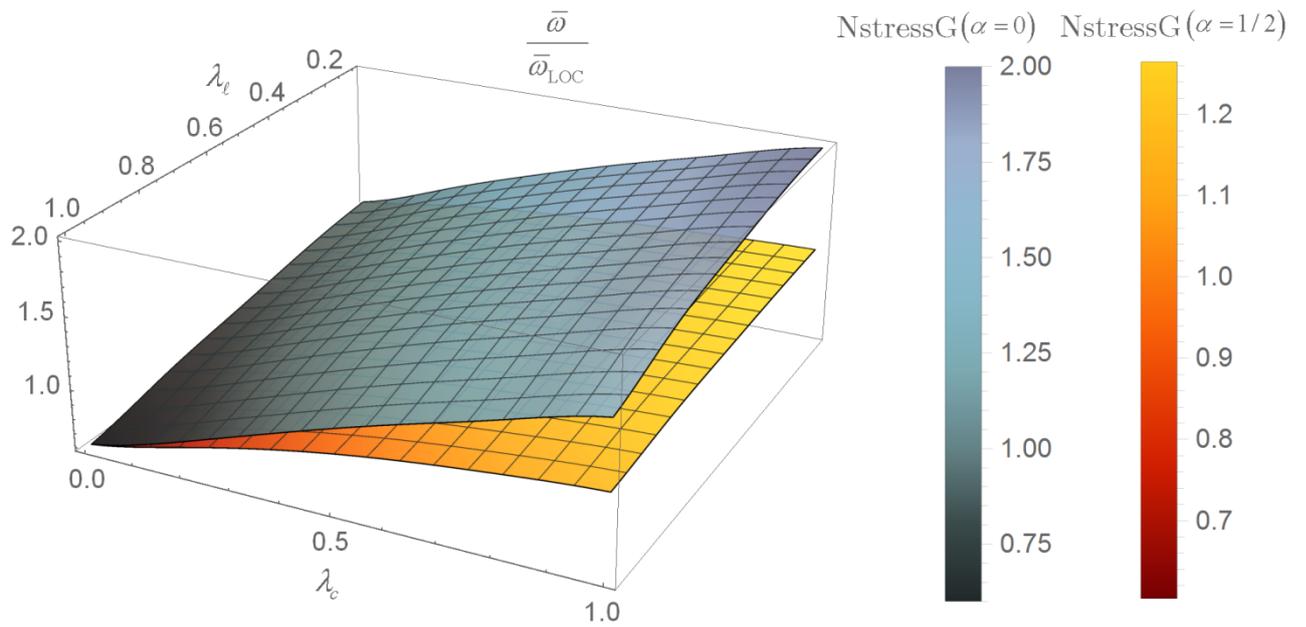

**Fig. 6.** Normalized flexural fundamental frequency of NstressG cantilever nano-beams

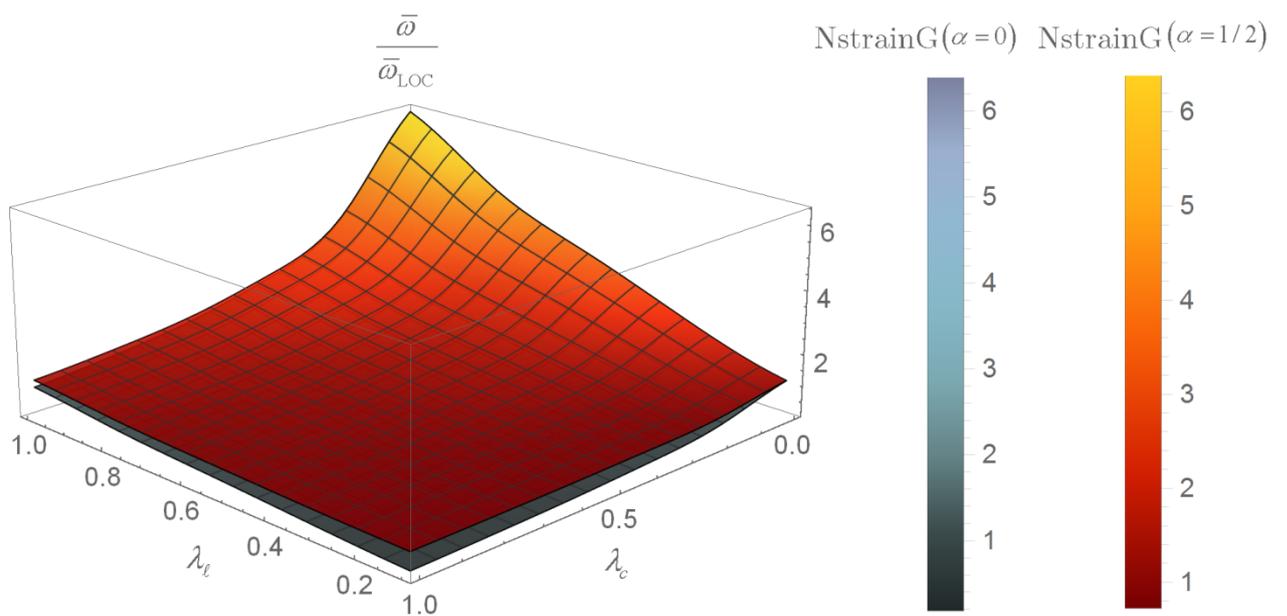

**Fig. 7.** Normalized flexural fundamental frequency of NstrainG fully-clamped nano-beams

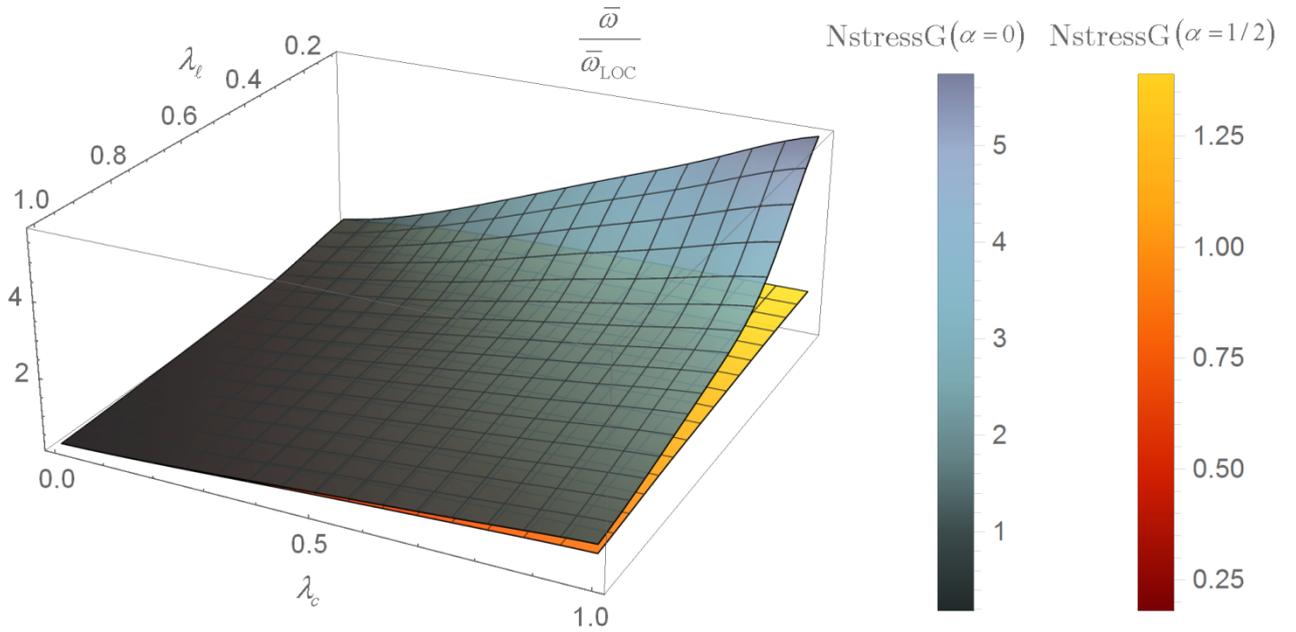

**Fig. 8.** Normalized flexural fundamental frequency of NstressG fully-clamped nano-beams

## 6. Closing remarks

The variational theory of Nonlocal strain and stress Gradient (NstrainG/NstressG) elasticity, conceived in (Barretta & Marotti de Sciarra, 2019) to study the static behavior of nano-structures, has been extended in this paper to predict dynamic responses of slender beams. The developed variational framework, equipped with appropriately selected test fields, has been employed to properly detect nonlocal gradient effects which significantly affect the elasto-dynamic behavior of most of advanced materials and smaller and smaller structures. The established integral laws have been then usefully transformed into equivalent differential conditions supplemented with suitable non-standard nonlocal gradient boundary conditions. The resulting nonlocal gradient elasticity formulation has been demonstrated to lead to well-posed continuum problems of dynamics of nanostructures and to provide, under special assumptions, the well-recognized models of nano-engineering. The procedure has been illustrated by examining the vibrational responses of cantilever and fully-clamped beams.

Axial and flexural frequency solutions have been analytically and numerically evaluated and graphically evidenced. NstrainG and NstressG elasticity strategies have been shown to be able to describe both softening and stiffening structural responses by appropriately driving the characteristic length-scale parameters. The paper provides therefore a reliable engineering approach to tackle modern dynamic problems in Nanoscience.

**Acknowledgment**

The financial support of the Italian Ministry for University and Research (P.R.I.N. National Grant 2017, Project code 2017J4EAYB; University of Naples Federico II Research Unit) is gratefully acknowledged.

**Table 1.** Normalized fundamental axial frequencies of NstrainG cantilever nano-beams

| | $\dfrac{\bar{\Omega}}{\bar{\Omega}_{\mathrm{LOC}}}$ | | | | | | | | | |
|---|---|---|---|---|---|---|---|---|---|---|
| | NstrainG $(\alpha = 0)$ | | | | | NstrainG $(\alpha = 1/2)$ | | | | |
| $\lambda_c$ | $\lambda_\ell = 0.1$ | $\lambda_\ell = 0.3$ | $\lambda_\ell = 0.5$ | $\lambda_\ell = 0.7$ | $\lambda_\ell = 1$ | $\lambda_\ell = 0.1$ | $\lambda_\ell = 0.3$ | $\lambda_\ell = 0.5$ | $\lambda_\ell = 0.7$ | $\lambda_\ell = 1$ |
| $0^+$ | 1.00881 | 1.04746 | 1.07204 | 1.08408 | 1.09229 | 1.00931 | 1.04795 | 1.07255 | 1.08461 | 1.09282 |
| 0.2 | 0.825677 | 0.880371 | 0.91814 | 0.93752 | 0.951042 | 0.938642 | 0.971574 | 1.00343 | 1.02323 | 1.03877 |
| 0.4 | 0.68401 | 0.752015 | 0.802117 | 0.829042 | 0.848321 | 0.885281 | 0.913787 | 0.946722 | 0.971764 | 0.99479 |
| 0.6 | 0.57862 | 0.652907 | 0.711743 | 0.744967 | 0.769457 | 0.849555 | 0.87186 | 0.901712 | 0.928252 | 0.9565 |
| 0.8 | 0.499123 | 0.575002 | 0.639306 | 0.677421 | 0.706371 | 0.825437 | 0.842294 | 0.867456 | 0.892693 | 0.923149 |
| 1.0 | 0.437734 | 0.512614 | 0.579944 | 0.621704 | 0.654384 | 0.808299 | 0.82111 | 0.841726 | 0.864351 | 0.894613 |

**Table 2.** Normalized fundamental axial frequencies of NstressG cantilever nano-beams

| | $\dfrac{\bar{\Omega}}{\bar{\Omega}_{\text{LOC}}}$ | | | | | | | | | |
|---|---|---|---|---|---|---|---|---|---|---|
| | NstressG $(\alpha = 0)$ | | | | | NstressG $(\alpha = 1/2)$ | | | | |
| $\lambda_c$ | $\lambda_\ell = 0.1$ | $\lambda_\ell = 0.3$ | $\lambda_\ell = 0.5$ | $\lambda_\ell = 0.7$ | $\lambda_\ell = 1$ | $\lambda_\ell = 0.1$ | $\lambda_\ell = 0.3$ | $\lambda_\ell = 0.5$ | $\lambda_\ell = 0.7$ | $\lambda_\ell = 1$ |
| $0^+$ | 0.993063 | 0.909824 | 0.791382 | 0.677167 | 0.540387 | 0.990588 | 0.908 | 0.790265 | 0.676525 | 0.540103 |
| 0.2 | 1.1342 | 1.05723 | 0.941077 | 0.822119 | 0.671564 | 1.05695 | 0.994419 | 0.896652 | 0.792509 | 0.655433 |
| 0.4 | 1.30962 | 1.24268 | 1.1357 | 1.01842 | 0.858455 | 1.122 | 1.07943 | 1.00751 | 0.923114 | 0.798816 |
| 0.6 | 1.47517 | 1.41578 | 1.31684 | 1.20263 | 1.03734 | 1.16966 | 1.13966 | 1.08636 | 1.0195 | 0.912237 |
| 0.8 | 1.62741 | 1.57361 | 1.4813 | 1.37066 | 1.20303 | 1.2046 | 1.18249 | 1.14195 | 1.08879 | 0.998065 |
| 1.0 | 1.76803 | 1.71854 | 1.63176 | 1.52474 | 1.35665 | 1.23101 | 1.2141 | 1.18244 | 1.13964 | 1.06325 |

**Table 3.** Normalized fundamental axial frequencies of NstrainG fully-clamped nano-beams

| | $\dfrac{\overline{\Omega}}{\overline{\Omega}_{\text{LOC}}}$ | | | | | | | | | |
|---|---|---|---|---|---|---|---|---|---|---|
| | NstrainG $(\alpha = 0)$ | | | | | NstrainG $(\alpha = 1/2)$ | | | | |
| $\lambda_c$ | $\lambda_\ell = 0.1$ | $\lambda_\ell = 0.3$ | $\lambda_\ell = 0.5$ | $\lambda_\ell = 0.7$ | $\lambda_\ell = 1$ | $\lambda_\ell = 0.1$ | $\lambda_\ell = 0.3$ | $\lambda_\ell = 0.5$ | $\lambda_\ell = 0.7$ | $\lambda_\ell = 1$ |
| $0^+$ | 1.04609 | 1.37139 | 1.85837 | 2.41097 | 3.29032 | 1.04705 | 1.37212 | 1.85891 | 2.41139 | 3.29062 |
| 0.2 | 0.702992 | 0.92573 | 1.2552 | 1.62873 | 2.22298 | 0.891416 | 1.07164 | 1.36566 | 1.71511 | 2.28695 |
| 0.4 | 0.508701 | 0.670681 | 0.909472 | 1.18015 | 1.61076 | 0.807949 | 0.918853 | 1.10594 | 1.33771 | 1.72967 |
| 0.6 | 0.394737 | 0.520155 | 0.705278 | 0.915153 | 1.24905 | 0.768124 | 0.840297 | 0.967124 | 1.13003 | 1.4145 |
| 0.8 | 0.321349 | 0.42311 | 0.573627 | 0.7443 | 1.01584 | 0.747484 | 0.797108 | 0.887309 | 1.00687 | 1.22195 |
| 1.0 | 0.27053 | 0.355938 | 0.482513 | 0.626059 | 0.854447 | 0.735658 | 0.771508 | 0.838258 | 0.929001 | 1.09659 |

**Table 4.** Normalized fundamental axial frequencies of NstressG fully-clamped nano-beams

| | $\dfrac{\bar{\Omega}}{\bar{\Omega}_{LOC}}$ | | | | | | | | | |
|---|---|---|---|---|---|---|---|---|---|---|
| | NstressG ($\alpha = 0$) | | | | | NstressG ($\alpha = 1/2$) | | | | |
| $\lambda_c$ | $\lambda_\ell = 0.1$ | $\lambda_\ell = 0.3$ | $\lambda_\ell = 0.5$ | $\lambda_\ell = 0.7$ | $\lambda_\ell = 1$ | $\lambda_\ell = 0.1$ | $\lambda_\ell = 0.3$ | $\lambda_\ell = 0.5$ | $\lambda_\ell = 0.7$ | $\lambda_\ell = 1$ |
| $0^+$ | 0.963918 | 0.734317 | 0.541138 | 0.41665 | 0.304945 | 0.959391 | 0.732568 | 0.54055 | 0.416422 | 0.304873 |
| 0.2 | 1.33352 | 1.01291 | 0.745053 | 0.573261 | 0.419511 | 1.11392 | 0.907395 | 0.700549 | 0.552435 | 0.411207 |
| 0.4 | 1.87819 | 1.43016 | 1.05406 | 0.811938 | 0.59467 | 1.23626 | 1.0792 | 0.887906 | 0.728493 | 0.559513 |
| 0.6 | 2.45556 | 1.87137 | 1.3801 | 1.06343 | 0.779043 | 1.30155 | 1.18712 | 1.02679 | 0.874625 | 0.69502 |
| 0.8 | 3.04322 | 2.32006 | 1.71144 | 1.31891 | 0.966297 | 1.33773 | 1.25351 | 1.12383 | 0.987753 | 0.810909 |
| 1.0 | 3.63536 | 2.772 | 2.04508 | 1.57614 | 1.15481 | 1.35929 | 1.29576 | 1.19149 | 1.07355 | 0.907297 |

**Table 5.** Normalized fundamental flexural frequencies of NstrainG cantilever nano-beams

| | $\dfrac{\overline{\omega}}{\overline{\omega}_{\text{LOC}}}$ | | | | | | | | | |
|---|---|---|---|---|---|---|---|---|---|---|
| | NstrainG $\left(\alpha = 0\right)$ | | | | | NstrainG $\left(\alpha = 1/2\right)$ | | | | |
| $\lambda_c$ | $\lambda_\ell = 0.1$ | $\lambda_\ell = 0.3$ | $\lambda_\ell = 0.5$ | $\lambda_\ell = 0.7$ | $\lambda_\ell = 1$ | $\lambda_\ell = 0.1$ | $\lambda_\ell = 0.3$ | $\lambda_\ell = 0.5$ | $\lambda_\ell = 0.7$ | $\lambda_\ell = 1$ |
| $0^+$ | 1.01778 | 1.10709 | 1.17522 | 1.21277 | 1.24021 | 1.01858 | 1.10769 | 1.17579 | 1.21335 | 1.24081 |
| 0.2 | 0.742578 | 0.842905 | 0.919966 | 0.964591 | 0.998419 | 0.908176 | 0.955825 | 1.01416 | 1.05529 | 1.09043 |
| 0.4 | 0.57857 | 0.678222 | 0.759563 | 0.809559 | 0.849128 | 0.852423 | 0.881443 | 0.924208 | 0.961697 | 0.999756 |
| 0.6 | 0.472341 | 0.566807 | 0.648583 | 0.701637 | 0.745402 | 0.819675 | 0.839178 | 0.870181 | 0.901559 | 0.93847 |
| 0.8 | 0.398493 | 0.486605 | 0.566748 | 0.621223 | 0.667894 | 0.798965 | 0.812676 | 0.835572 | 0.861027 | 0.894649 |
| 1.0 | 0.344364 | 0.426181 | 0.503698 | 0.558531 | 0.607142 | 0.784817 | 0.794818 | 0.812133 | 0.832659 | 0.862289 |

**Table 6.** Normalized fundamental flexural frequencies of NstressG cantilever nano-beams

| | $\dfrac{\overline{\omega}}{\overline{\omega}_{\mathrm{LOC}}}$ | | | | | | | | | |
|---|---|---|---|---|---|---|---|---|---|---|
| | NstressG $(\alpha = 0)$ | | | | | NstressG $(\alpha = 1/2)$ | | | | |
| $\lambda_c$ | $\lambda_\ell = 0.1$ | $\lambda_\ell = 0.3$ | $\lambda_\ell = 0.5$ | $\lambda_\ell = 0.7$ | $\lambda_\ell = 1$ | $\lambda_\ell = 0.1$ | $\lambda_\ell = 0.3$ | $\lambda_\ell = 0.5$ | $\lambda_\ell = 0.7$ | $\lambda_\ell = 1$ |
| $0^+$ | 0.991445 | 0.924749 | 0.826851 | 0.728411 | 0.604041 | 0.990958 | 0.924353 | 0.826564 | 0.72821 | 0.60392 |
| 0.2 | 1.21717 | 1.14936 | 1.0453 | 0.935494 | 0.789909 | 1.08962 | 1.04005 | 0.960488 | 0.872263 | 0.749191 |
| 0.4 | 1.44445 | 1.38385 | 1.28526 | 1.17415 | 1.01623 | 1.16113 | 1.12886 | 1.07277 | 1.00424 | 0.89737 |
| 0.6 | 1.64847 | 1.59405 | 1.50187 | 1.39279 | 1.22902 | 1.20831 | 1.18632 | 1.14632 | 1.09433 | 1.00644 |
| 0.8 | 1.83235 | 1.78272 | 1.69628 | 1.59034 | 1.42441 | 1.24099 | 1.2252 | 1.19568 | 1.15578 | 1.08447 |
| 1.0 | 2.00041 | 1.95455 | 1.87306 | 1.77054 | 1.60453 | 1.26483 | 1.25298 | 1.23045 | 1.19921 | 1.14118 |

**Table 7.** Normalized fundamental flexural frequencies of NstrainG fully-clamped nano-beams

| | $\dfrac{\overline{\omega}}{\overline{\omega}_{\mathrm{LOC}}}$ | | | | | | | | | |
|---|---|---|---|---|---|---|---|---|---|---|
| | NstrainG $\left(\alpha=0\right)$ | | | | | NstrainG $\left(\alpha=1/2\right)$ | | | | |
| $\lambda_c$ | $\lambda_\ell=0.1$ | $\lambda_\ell=0.3$ | $\lambda_\ell=0.5$ | $\lambda_\ell=0.7$ | $\lambda_\ell=1$ | $\lambda_\ell=0.1$ | $\lambda_\ell=0.3$ | $\lambda_\ell=0.5$ | $\lambda_\ell=0.7$ | $\lambda_\ell=1$ |
| $0^+$ | 1.18839 | 2.15561 | 3.33222 | 4.55943 | 6.4318 | 1.18946 | 2.15615 | 3.33257 | 4.55968 | 6.43197 |
| 0.2 | 0.628169 | 1.15636 | 1.79059 | 2.45124 | 3.45877 | 0.875322 | 1.30136 | 1.88678 | 2.52219 | 3.50935 |
| 0.4 | 0.393436 | 0.725658 | 1.12389 | 1.53864 | 2.17114 | 0.777495 | 0.985618 | 1.30649 | 1.67655 | 2.2709 |
| 0.6 | 0.28134 | 0.519051 | 0.803918 | 1.1006 | 1.55303 | 0.743728 | 0.860957 | 1.05722 | 1.29713 | 1.69795 |
| 0.8 | 0.217621 | 0.401504 | 0.621857 | 0.851348 | 1.20132 | 0.729158 | 0.802685 | 0.932529 | 1.09891 | 1.38778 |
| 1.0 | 0.176963 | 0.326484 | 0.505661 | 0.69227 | 0.976846 | 0.72173 | 0.771612 | 0.862772 | 0.983807 | 1.20118 |

**Table 8.** Normalized fundamental flexural frequencies of NstressG fully-clamped nano-beams

| | $\dfrac{\overline{\omega}}{\overline{\omega}_{\mathrm{LOC}}}$ | | | | | | | | | |
|---|---|---|---|---|---|---|---|---|---|---|
| | NstressG $(\alpha = 0)$ | | | | | NstressG $(\alpha = 1/2)$ | | | | |
| $\lambda_c$ | $\lambda_\ell = 0.1$ | $\lambda_\ell = 0.3$ | $\lambda_\ell = 0.5$ | $\lambda_\ell = 0.7$ | $\lambda_\ell = 1$ | $\lambda_\ell = 0.1$ | $\lambda_\ell = 0.3$ | $\lambda_\ell = 0.5$ | $\lambda_\ell = 0.7$ | $\lambda_\ell = 1$ |
| $0^+$ | 0.891651 | 0.525222 | 0.346189 | 0.254668 | 0.181217 | 0.884156 | 0.523144 | 0.345285 | 0.254125 | 0.180881 |
| 0.2 | 1.58127 | 0.898224 | 0.58545 | 0.428962 | 0.304529 | 1.14531 | 0.779381 | 0.541257 | 0.406088 | 0.292398 |
| 0.4 | 2.57577 | 1.38724 | 0.937129 | 0.685743 | 0.486459 | 1.28946 | 1.02374 | 0.776246 | 0.606595 | 0.448734 |
| 0.6 | 3.6194 | 2.0164 | 1.30727 | 0.95609 | 0.678036 | 1.34633 | 1.1675 | 0.954526 | 0.780061 | 0.59726 |
| 0.8 | 4.67784 | 2.59796 | 1.68291 | 1.23048 | 0.872488 | 1.37237 | 1.24944 | 1.07846 | 0.917226 | 0.727892 |
| 1.0 | 5.74258 | 3.18308 | 2.06089 | 1.50658 | 1.06816 | 1.38608 | 1.29819 | 1.1632 | 1.02187 | 0.838555 |